\documentclass[a4paper,12pt]{article}

\usepackage[left=2.5cm,right=2.5cm,top=2.5cm,bottom=2.5cm]{geometry}
\usepackage{pslatex}
\usepackage{textpos}
\usepackage{caption}
\usepackage{listings}
\usepackage{color}
\usepackage{textcomp}
\usepackage{graphicx,xcolor}
\usepackage{epsfig}
\usepackage{epstopdf}
\usepackage{pgf}
\usepackage{amsmath,amssymb,amsfonts,amsthm,breqn,amsbsy}
\usepackage{longtable}
\usepackage{verbatim}
\usepackage{float}
\usepackage{appendix}
\usepackage{array}
\usepackage{varwidth}
\usepackage{multirow}
\usepackage{tabularx}
\usepackage{framed}
\usepackage{lipsum}
\usepackage[round]{natbib}
\usepackage{hyperref}
\usepackage{breakurl}

\usepackage{breakcites}
\usepackage{longtable}
\usepackage[utf8]{inputenc}
\usepackage{booktabs}
\usepackage{lscape}
\usepackage{rotating}
\usepackage{fancyvrb}
\usepackage{setspace}
\usepackage[T1]{fontenc}
\usepackage{indentfirst}
\usepackage{lineno}
\usepackage{mathtools}
\usepackage{bigints}
\usepackage{url}
\usepackage{bm,bbm}
\usepackage[normalem]{ulem}
 \usepackage{microtype} 
 \newcommand{\pkg}[1]{{\fontseries{b}\selectfont #1}}
\let\proglang=\textsf

\title{Bayesian analysis of diffusion-driven multi-type epidemic models with application to COVID-19}
\author{Lampros Bouranis$^{\ast,1}$, Nikolaos Demiris$^1$, \\ Konstantinos Kalogeropoulos$^2$, Ioannis Ntzoufras$^1$}

\makeatletter
\renewcommand\@date{{%
  \large\centering
  $^{1}$\textit{Department of Statistics, Athens University of Economics and Business, Athens, Greece}
  \par
  \vspace{0.5ex}
  $^{2}$\textit{Department of Statistics, The London School of Economics and Political Science, London, United kingdom}
  \par
  \vspace{0.5ex}
  $^{\ast}$ bouranis@aueb.gr
}}
\makeatother

\begin{document}

\maketitle

\begin{abstract}
We consider a flexible Bayesian evidence synthesis approach to model the age-specific transmission dynamics of COVID-19 based on daily mortality counts. 
The temporal evolution of transmission rates in populations containing multiple types of individual is reconstructed via an appropriate dimension-reduction formulation 
driven by independent diffusion processes. A suitably tailored compartmental model is used to learn the latent counts of infection, accounting for fluctuations in 
transmission influenced by public health interventions and changes in human behaviour. The model is fitted to freely available COVID-19 data sources from the UK, 
Greece and Austria and validated using a large-scale prevalence survey in England. In particular, we demonstrate how model expansion can facilitate evidence reconciliation at a latent level. 
The code implementing this work is made freely available via the \pkg{Bernadette} \proglang{R} package.
\end{abstract}

\noindent%


\section{Introduction}\label{section:intro}
The COVID-19 outbreak caused by the Severe acute respiratory syndrome coronavirus 2 (SARS-CoV-2), has led to developments in Bayesian infectious disease modeling, allowing modelers to assess the impact of mitigation strategies on transmission and quantify the burden of the pandemic \citep{flaxman, monod}. 
The highly transmissible nature of COVID-19 has proven to be a challenge for health systems worldwide, particularly in the acute phase of the pandemic, pre-vaccination. A number of large-scale prevalence studies aiming to estimate the actual number of infections have found severe under-ascertainment \citep{ward_react2}, which has been varying in time and across countries. 
The level of under-ascertainment depends on the national testing and tracing policies, the testing capacities and the impact of false positives under different regimes. Especially at the early stages of the pandemic, those tested were typically more likely to have been hospitalised or were at higher risk of infection or death, leading to only a proportion of infections being detected and reported \citep{li}. Consequently, methods that rely on reported counts of infections are expected to yield biases in the inferred rates of transmission.

This work is driven by the challenges of under-ascertainment of COVID-19 infections and the presence of heterogeneity in type, relevance, and granularity of the data. Building upon the approach by \cite{flaxman} and its extension to multiple age groups by \cite{monod}, we propose a Bayesian evidence synthesis approach for the analysis of COVID-19 data, with the aim to: (i) infer the true number of infections using age-stratified daily COVID-19 attributable mortality counts; (ii) learn the age-specific transmission rates; (iii) reconstruct the epidemic drivers from publicly available data sources. 

\cite{chatzilena} extend the work of \cite{dureau} allowing for indirectly observed infections. 
We extend this framework by developing an age-stratified stochastic transmission model with distinct transmissibility parameters between age groups accounting for the presence of social structures. In particular, we target the time-varying transmission rate matrix process, the dimension of which increases quadratically with the number of age groups, and offer a dimension-reduction formulation projecting to latent diffusion processes. This formulation possesses several desirable characteristics in capturing the multiple group disease transmission mechanism: (i) it is based on a natural decomposition into the underlying biological and social components; (ii) it allows for further evidence synthesis utilising information from contact surveys; (iii) it is driven by a potentially non-scalar diffusion process to adequately capture the temporal evolution of the transmission rates, as well as extrinsic environmental factors such as non-pharmaceutical interventions (NPIs) and climatic changes \citep{cazelles, chatzilena, ghosh}; (iv) it facilitates model determination at a latent level, performed here by appropriate model expansion.

In this paper we offer a modeling framework which is suitable for evidence synthesis. Capturing the time-varying nature of the different mechanisms involved in disease propagation is a complex task \citep{birrell2}. Modeling disease transmission as a stochastic process offers a flexible Bayesian non-parametric description of the infection process at the population level that imposes minimal restrictions to the transmission rate trajectories. The proposed model is agnostic about the timing of specific interventions and their impact on transmission using less information than related approaches \citep{knock, birrell}. We demonstrate how model assessment can be conducted by exploring the latent components of the proposed model, and provide an easy and efficient implementation for researchers with access to freely available data. We use the Stan \citep{carpenter2017stan} open-source software's Hamiltonian Monte Carlo algorithm \citep{betancourt} while particle Markov chain Monte Carlo (MCMC) may also be applied \citep{dureau}.
We do not assume availability of healthcare surveillance data sources such as serological survey data (antibody testing), hospital admissions, and ICU occupancy but allow for the integration of multiple, publicly available, data sources: age-stratified daily COVID-19 attributable mortality counts, contact surveys and the age distribution of the population. 
The contact data are used to delineate potential identifiability issues \citep{britton} at the unobserved infection rate level where those rates are decomposed in their social and biological component.

The outline of the paper is as follows. Section \ref{section:methods} provides an overview of the multiple data sources and a presentation of the components of the Bayesian hierarchical model. 
In Section \ref{section:application} we present the results of the empirical analysis for Greece and discuss the effect of model expansion on resolving prior-data conflict at a latent level. The proposed model is validated using estimates from a large-scale prevalence survey in England. Results for Austria are presented in the Supplementary Material.
We conclude the paper in Section \ref{section:discussion} with final remarks. 

\section{Materials and Methods}\label{section:methods}

\subsection{Data sources}\label{section:data}
Our model is driven by the characteristics of SARS-CoV-2 transmission and applied to observations of the epidemic in Greece, Austria and England. 
For each of these countries, we collected three types of data related to SARS-CoV-2 spread: (i) the age distribution of reported deaths; (ii) the age distribution of all reported infections; (iii) the age distribution of the general population for each country, $\mathbb{N}$. 
The study period for Austria ranges from 2020-08-31 to 2021-03-28 (30 weeks). The study period for Greece ranges from 2020-09-01 to 2021-03-29 (30 weeks). 
For model validation, we analyse the period 2020-03-02 to 2020-09-27 (30 weeks) in England.

We adopted the country-specific synthetic contact matrices of \cite{prem1}, which have been constructed upon adjusted contact patterns, based on national demographic and socioeconomic characteristics. In the absence of a synthetic contact matrix for England we used the one created for the United Kingdom, a reasonable assumption due to the population overlap. The elements of these matrices represent the daily average numbers of contacts between age groups.

The severity of the disease is incorporated in the model via the age-stratified infection fatality rate (IFR), informed by the REal-time Assessment of Community Transmission-2 (REACT-2) national prevalence study of SARS-CoV-2, of over 100,000 adults in England \citep{ward_react2}.
Detailed references for each data source are presented in Section S1 of the Supplementary Material.

\subsection{Modeling framework}\label{section:mod_framework}

\subsubsection{Evidence synthesis}\label{section:evidence_synthesis}
\begin{figure}[ht]
\centering
\includegraphics[scale=0.45]{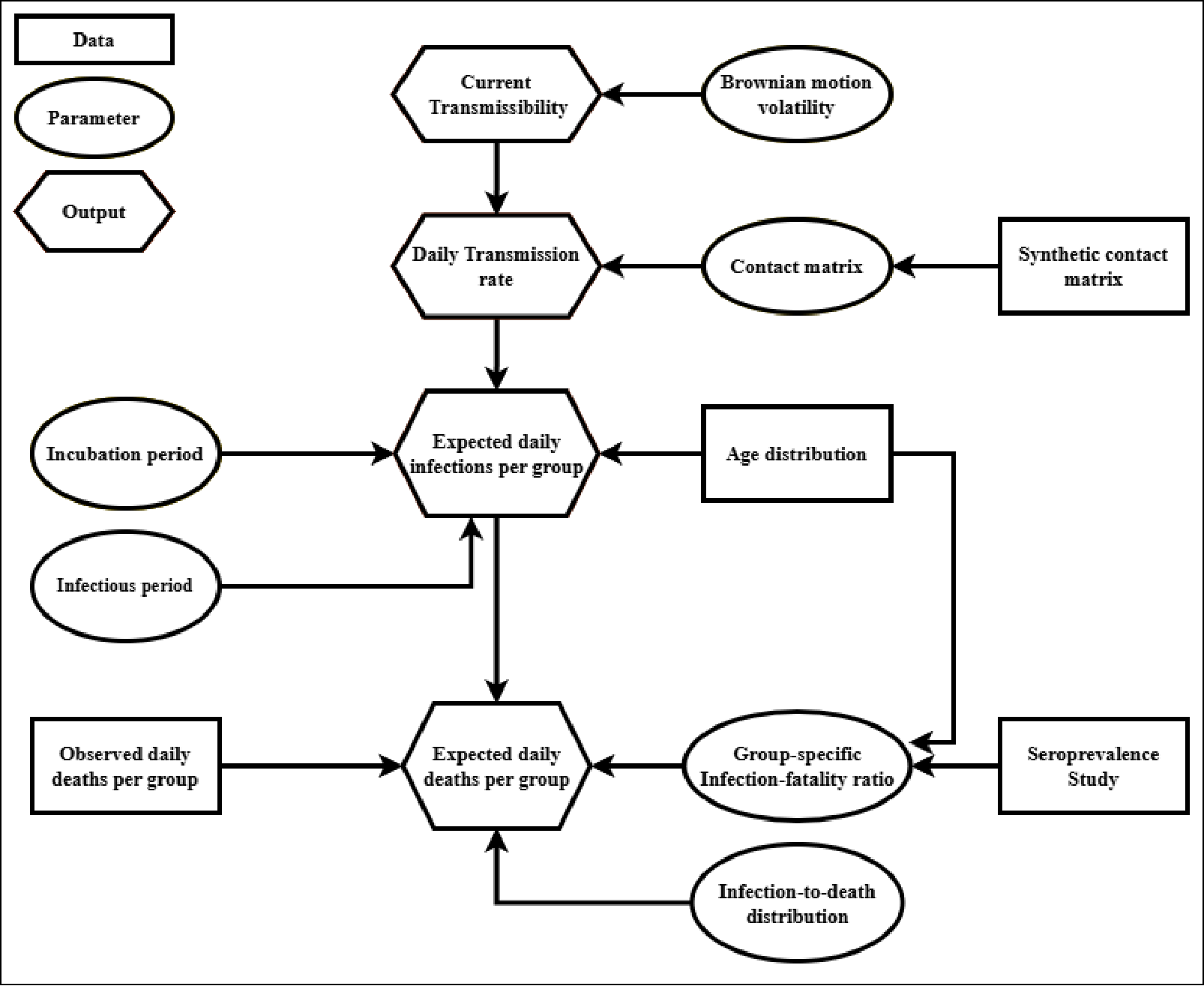}
\caption{Directed acyclic graph describing the functional relationships between data sources (rectangles), modeled outputs (ovals) and parameters (hexagons).}
\label{fig:functional_relationships}
\end{figure}

The aforementioned data streams and expert knowledge were integrated into a coherent modeling framework via a Bayesian evidence synthesis approach \citep{deangelis}. Following the works of \cite{flaxman}, \cite{monod} and \cite{chatzilena}, the modeling process is decomposed into the latent epidemic process and the observation model to facilitate clarity.
Figure \ref{fig:functional_relationships} shows the functional relationships between data sources, modeled outputs and parameters in this study. 
The components of the Bayesian hierarchical model are laid out in the remainder of Section \ref{section:mod_framework}. All parts of the modeling process are country-specific, therefore a country index has been omitted in the following for brevity.

\subsubsection{Diffusion-driven multi-type transmission process}\label{section:trans_model}
\begin{sloppypar}
The transmission of COVID-19 is modeled through an age-stratified stochastic Susceptible-Exposed-Infectious-Removed (SEIR) compartmental model \citep{anderson} with Erlang-distributed latent and infectious periods. 
In particular, we introduce Erlang-distributed stage durations in the Exposed and Infectious compartments and relax the mathematically convenient but unrealistic assumption of exponential stage durations \citep{watson, lloyd} by assuming that each of the Exposed and Infectious compartments are defined by two stages, with the same rate of loss of latency ($\tau$) and infectiousness ($\gamma$) in both stages.
Removed individuals are assumed to be immune to reinfection for at least the duration of the study period.
\end{sloppypar}

The population is stratified into $\alpha \in \{1,\ldots,A\}$ age groups and the total size of the age group is denoted by $\mathbb{N}_{\alpha} = S_t^{\alpha} + E_{1,t}^{\alpha} + E_{2,t}^{\alpha} + I_{1,t}^{\alpha} + I_{2,t}^{\alpha} + R_t^{\alpha}$, where $S_t^{\alpha}$ represents the number of susceptible, $E_t^{\alpha} = \sum_{j=1}^{2}E_{j,t}^{\alpha}$ represents the number of exposed but not yet infectious, $I_t^{\alpha} = \sum_{j=1}^{2}I_{j,t}^{\alpha}$ is the number of infected and $R_t^{\alpha}$ is the number of removed individuals at time $t$ at age group $\alpha$. 
The number of individuals in each compartment is scaled by the total population $\mathbb{N} = \sum_{\alpha = 1}^{A}\mathbb{N}_{\alpha} $, so that the sum of all compartments equals to one \citep{riou2}.
The latent epidemic process is expressed by the following continuous-time non-linear system of ordinary differential equations (ODEs) \citep{kermack}
\begin{equation}
\begin{cases}\label{eq:ODE_system} 
\frac{dS^{\alpha}_{t}}{dt}   & = -\lambda_{\alpha}(t) S^{\alpha}_{t},
\\
\frac{dE^{\alpha}_{1,t}}{dt} & = \lambda_{\alpha}(t) S^{\alpha}_{t} - \tau E^{\alpha}_{1,t},
\\
\frac{dE^{\alpha}_{2,t}}{dt} & = \tau \left( E^{\alpha}_{1,t} - E^{\alpha}_{2,t}\right),
\\
\frac{dI^{\alpha}_{1,t}}{dt} & = \tau E^{\alpha}_{2,t} - \gamma I^{\alpha}_{1,t},
\\
\frac{dI^{\alpha}_{2,t}}{dt} & = \gamma \left( I^{\alpha}_{1,t} - I^{\alpha}_{2,t}\right),
\\
\frac{dR^{\alpha}_{t}}{dt}   & = \gamma I^{\alpha}_{2,t}
\end{cases}
\end{equation}
where the mean latent and infectious periods are $d_E = \frac{2}{\tau}$, $d_I = \frac{2}{\gamma}$, respectively, and time is approximated in discrete steps of one day, following the granularity of the mortality data. 

The age-stratified transmission dynamics depend, among others, on the initial conditions of the system of non-linear ODEs. The date of simulation start $t_0$ is set on the day before the date of the first data point $t_1$ (Supplementary Material, Table S2). At $t_0$, most individuals are assumed to be susceptible and are distributed across age groups according to the population age distribution $\mathbb{N}_{\alpha}$ for $\alpha \in \{1,\ldots,A\}$. The number of people in the Exposed compartments at $t_0$ is controlled by a parameter $\rho$ (Supplementary Material, Table S2), and also distributed according to $\mathbb{N}_{\alpha}$ for $\alpha \in \{1,\ldots,A\}$. Numbers in the other compartments are set to zero \citep{riou1}.

The number of new infections in age group $\alpha$ at time $t$ is governed by
\begin{equation}\label{eq:new_cases}
\Delta^{\text{infec}}_{t, \alpha} = \int_{t-1}^{t} \tau E^{\alpha}_{2,s} ds.
\end{equation}
The time-dependent force of infection $\lambda_{\alpha}(t)$ for age group $\alpha \in \{1,\ldots,A\}$ is expressed as
\begin{equation*}
\lambda_{\alpha}(t) = \sum_{\alpha'=1}^{A}\left[ m_{\alpha\alpha'}(t) \frac{\left(I^{\alpha'}_{1,t} + I^{\alpha'}_{2,t}\right)}{\mathbb{N}_{\alpha'}}\right],
\end{equation*}
which is a function of the proportion of infectious individuals in each age group $\alpha' \in \{1,\ldots,A\}$, via the compartments $I^{\alpha'}_{1,t}$, $I^{\alpha'}_{2,t}$ divided by the total size of the age group $\mathbb{N}_{\alpha'}$, and the time-varying person-to-person transmission rate from group $\alpha$ to group $\alpha'$, $m_{\alpha\alpha'}(t)$.
We parameterize the transmission rate between different age groups $(\alpha,\alpha') \in \{1,\ldots,A\}^2$ by
\begin{equation}\label{eq:transmission_matrix_process}
m_{\alpha\alpha'}(t) = 
\beta^{\alpha\alpha'}_{t} 
\cdot
C_{\alpha\alpha'},
\end{equation}
breaking down the transmission rate matrix into its biological and social components, considered, for instance, in \cite{knock} and \cite{baguelin}: the social component is represented by the average number of contacts between individuals of age group $\alpha$ and age group $\alpha'$ via the contact matrix element $C_{\alpha\alpha'}$; 
$\beta^{\alpha\alpha'}_{t}$ is the time-varying transmissibility of the virus, expressed by the probability that a contact between an infectious person in age group $\alpha$ and a susceptible person in age group $\alpha'$ leads to transmission at time $t$.

The formulation below may be viewed as a stochastic extension to the deterministic multi-type SEIR model, using diffusion processes for the coefficients $\beta^{\alpha\alpha'}_{t}$ in \eqref{eq:transmission_matrix_process}, driven by independent Brownian motions
\begin{equation}
\begin{cases}\label{eq:Euler_BM} 
\beta^{\alpha\alpha'}_{t} & = \exp(x^{\alpha\alpha'}_{t}), 
\\
x^{\alpha\alpha'}_{t} \mid x^{\alpha\alpha'}_{t - 1}, \sigma_{\alpha\alpha'} & \sim 
\operatorname{N} \Big( x^{\alpha\alpha'}_{t - 1}, \,\, \sigma_{\alpha\alpha'}^2 \Big), 
\\
dx^{\alpha\alpha'}_{t} & = \sigma_{\alpha\alpha'} \,  dW^{\alpha\alpha'}_{t},
\end{cases}
\end{equation}
with volatilities $\sigma_{\alpha\alpha'}$, where $W^{\alpha\alpha'}$ is a standard Brownian motion.
The volatility $\sigma_{\alpha\alpha'}$ plays the role of the regularizing factor: higher values of $\sigma_{\alpha\alpha'}$ lead to greater changes in $\beta^{\alpha\alpha'}_{t}$. 
The exponential transformation avoids negative values which have no biological meaning \citep{dureau, cazelles}. 
A major advantage of considering a diffusion process for modeling $\beta^{\alpha\alpha'}_{t}$ is its ability to capture and quantify the randomness of the underlying transmission dynamics, which is particularly useful when the dynamics are not completely understood. 
The diffusion process accounts for fluctuations in transmission that are influenced by non-modeled phenomena, such as new variants, mask-wearing propensity, etc. 
The diffusion process also allows for capturing the effect of unknown extrinsic factors on the age-stratified force of infection,
for monitoring of the temporal evolution of the age-specific transmission rate without the implicit inclusion of external variables and for tackling non-stationarity in the data.

We propose a diffusion-driven multi-type latent transmission model which assigns independent Brownian motions to $\log(\beta^{11}_{t}), \log(\beta^{22}_{t}), \ldots, \log(\beta^{AA}_{t})$ with respective age-stratified volatility parameters $\sigma_{\alpha}, \alpha \in \{1,\ldots,A\}$, for reasons of parsimony and interpretability. 
The contact matrix is scaled by the age-stratified transmissibility in order to obtain the transmission rate matrix process
\begin{equation}\label{eq:transm_rate_indep_bm}
m^{\text{MBM}}_{\alpha\alpha'}(t) = \beta^{\alpha\alpha'}_{t}
\cdot
C_{\alpha\alpha'}
\equiv 
\beta^{\alpha\alpha}_{t}
\cdot
C_{\alpha\alpha'},
\end{equation}
under the assumption $\beta^{\alpha\alpha'}_t \equiv \beta^{\alpha\alpha}_t, \alpha \neq \alpha'$. 
Henceforth, we term the model in \eqref{eq:transm_rate_indep_bm} by MBM (standing for Multi Brownian Motion).
The MBM model can offer dimension reduction from all the elements of the transmission rate matrix process to the Brownian motions $\beta_t^{\alpha\alpha}$ for $\alpha \in \{1,\ldots,A\}$. 
An appealing feature of this representation of the transmission rate matrix process is that it separates the contact matrix which reflects social components of the model and can be informed from survey data.

A prior distribution is imposed on the contact matrix elements $C_{\alpha\alpha^{\prime}}$, to account for the uncertainty of the contact structure while estimating disease transmission. We provide details on the formulation of the contact matrix in the Supplementary Material.

Further reduction in the dimension of the transmission rate matrix process can be achieved by the formulation
\begin{equation}\label{eq:transm_rate_commom_bm}
m^{\text{SBM}}_{\alpha\alpha'}(t) = 
\beta^{\alpha\alpha'}_{t} 
\cdot
C_{\alpha\alpha'}
\equiv 
\beta_{t} 
\cdot 
C_{\alpha\alpha'},
\end{equation}
under the assumption $\beta^{\alpha\alpha'}_t \equiv \beta_{t}$, where $\beta_{t}$ is the time-varying transmissibility of the virus, the probability that a contact between an infectious person and a susceptible person leads to transmission at time $t$.
Henceforth, we term the model in \eqref{eq:transm_rate_commom_bm} by SBM (standing for Single Brownian Motion), which represents a nested model to MBM.

In related work, \cite{birrell} estimated the transmission rate between different age groups with a variation of the SBM model, considering time-dependent (weekly-varying) contact matrices while \cite{ghosh} adapted the model of \cite{birrell} and expressed virus transmissibility by a discretised path-wise approximation of a Brownian motion.
\cite{knock} considered a time-independent contact matrix while the transmission rate between different age groups was parameterised as in \eqref{eq:transm_rate_commom_bm} and the virus transmissibility was assumed to be piecewise linear with multiple change points corresponding to major announcements and changes in COVID-19 related policy.
\cite{tan} and \cite{bouman} parameterised the age-stratified transmission rate as in \eqref{eq:transm_rate_indep_bm}, considering B-splines for the age-stratified virus transmissibility.

\subsubsection{Observation process}\label{section:obs_model}
Denote the number of observed deaths on day $t = 1, \ldots, T$ in age group $\alpha \in \{1,\ldots,A\}$ by $y_{t,\alpha}$. 
A given infection may lead to observation events (i.e deaths) in the future. 
A link between $y_{t,\alpha}$ and the expected number of new age-stratified infections is established via the function
\begin{equation*}
d_{t,\alpha} = \mathbb{E}[y_{t,\alpha}] =\text{IFR}_{\alpha} \times \sum_{s = 1}^{t-1}h_{t-s} \Delta^{\text{infec}}_{s, \alpha}
\end{equation*}
on the new expected age-stratified mortality counts, $d_{t,\alpha}$, the age-stratified infection fatality rate, $\text{IFR}_{\alpha}$, and the infection-to-death distribution $h$, where $h_s$ gives the probability the death occurs on the $s^{th}$ day after infection, is assumed to be Gamma distributed with mean 24.2 days and coefficient of variation 0.39 \citep{flaxman, monod}. 
Denoting the distribution of a Gamma random variable with shape $\zeta$ and scale $b$ by $\operatorname{Gamma}(\zeta, b)$, the infection-to-death distribution follows
\begin{equation}\label{eq:itd}
h \sim \operatorname{Gamma}(6.57, 3.68).
\end{equation}
In Section \ref{section:application}, the estimates of $\text{IFR}_{\alpha}$ from the REACT-2 study are weighted by the age distribution of the country-specific population to inform estimation of $d_{t,\alpha}$. 
We allow for over-dispersion in the observation processes to account for noise in the underlying data streams, for example due to day-of-week effects on data collection \citep{stoner, knock}, and link $d_{t,\alpha}$ to $y_{t,\alpha}$ through an over-dispersed count model \citep{riou1, birrell, seaman}
\begin{equation}\label{eq:negbin}
y_{t,\alpha}\mid \theta \sim \operatorname{NegBin}\big(d_{t,\alpha}, \, \xi_{t,\alpha}\mid \theta\big),
\end{equation}
where $\xi_{t,\alpha} = \frac{d_{t,\alpha}}{\phi}$, such that $\mathbb{V}[y_{t,\alpha}] = d_{t,\alpha}(1+\phi)$. 
The likelihood of the observed deaths is given by
\begin{equation*}\label{eq:loglike}
p(y\mid \theta) = \prod_{t=1}^{T}\prod_{\alpha=1}^{A}\text{NegBin}\big(y_{t,\alpha} \mid d_{t,\alpha}, \, \xi_{t,\alpha}, \, \theta \big),
\end{equation*}
for a set of parameters $\theta$, where $y \in \mathbb{R}^{T \times A}_{0,+}$ are the surveillance data on deaths for all time-points and age groups and the parameter vector corresponds to either $\theta^{SBM} = \big(x_{0}, \, x_{1:T}, \, C_{\alpha\alpha'}, \, \sigma, \, \rho, \, \phi \big)$ or $\theta^{MBM} = \big(x^{\alpha\alpha}_{0}, \,x^{\alpha\alpha}_{1:T}, \, C_{\alpha\alpha'},\, \sigma_{\alpha},\, \rho, \,\phi\big)$ for $(\alpha,\alpha') \in \{1,\ldots,A\}^2$.
The observation process in this work does not additionally model the age-stratified infection counts, for reasons discussed in Section \ref{section:intro}.

\subsection{Parameter estimation}\label{section:estimation}
The posterior of interest reads
\begin{align*}
\pi(\theta\mid  y) = \frac{p(y\mid \theta)p(\theta)}{\pi(y)},
\end{align*}
where $p(\theta)$ is the prior distribution on the parameter vector $\theta$, listed in Supplementary Material, Table S2. The posterior distributions of the SBM and MBM transmission models are 

\begin{align*} 
\pi \big(\theta^{SBM} \big|  y\big) & \propto \prod_{t=1}^{T}\prod_{\alpha=1}^{A}\text{NegBin}\left(y_{t,\alpha}\mid d_{t,\alpha}, \xi_{t,\alpha}, \theta^{SBM}\right) \\
& \mathrel{\phantom{=}}
\times
\prod_{\alpha=1}^{A}\prod_{\alpha^{\prime}=1}^{A}p( C_{\alpha\alpha'})\prod_{t=1}^{T} p(x_t\mid x_{t-1}, \sigma) p(x_0) p(\sigma) p(\rho) p(\phi) \\
\pi(\theta^{MBM}\mid y) & \propto 
\prod_{t=1}^{T}\prod_{\alpha=1}^{A}\text{NegBin}\left(y_{t,\alpha}\mid d_{t,\alpha}, \xi_{t,\alpha}, \theta^{MBM}\right) \\
& \mathrel{\phantom{=}}
\times
\prod_{\alpha = 1}^{A}\prod_{\alpha^{\prime} = 1}^{A}p( C_{\alpha\alpha'})
\prod_{t=1}^{T}\prod_{\alpha = 1}^{A} p(x^{\alpha\alpha}_t\mid x^{\alpha\alpha}_{t-1}, \sigma_{\alpha}) \prod_{\alpha = 1}^{A}p(x_0^{\alpha\alpha})\prod_{\alpha = 1}^{A}p(\sigma_{\alpha}) 
p(\rho) p(\phi).
\end{align*}

Bayesian estimation of the parameters of the SBM and MBM models was performed using MCMC due to intractability of $\pi(\theta^{SBM}\mid y)$ and $\pi(\theta^{MBM}\mid y)$.
Computational details are provided in the Supplementary Material.
In the following we discuss the challenging aspects of parameter estimation and present inference approximations used in the implementation in order to improve mixing and computational efficiency.

Estimation of the new daily infection counts in \eqref{eq:new_cases} requires the solution of the non-linear system of ODEs in \eqref{eq:ODE_system} coupled with the stochastic differential equation in \eqref{eq:Euler_BM}, which together are a hypo-elliptic diffusion which is intractable. 
We address the intractability of the hypo-elliptic diffusion by adopting the data augmentation framework of \cite{dureau} as a means to infer the latent sample path $x$ of the diffusion which is an infinite-dimensional object \citep{kalogeropoulos} and is indirectly observed through the time evolution of the disease states.
The task of inference for the transmission rate(s) via diffusions is particularly challenging; modeling its time evolution at a more granular time-scale (i.e. at the time-points of the observations) would naturally increase the cost of the computational effort; 
to reduce the dimensionality of the stochastic process we split the study period into $k = 1, \ldots, K$ weeks and denote by $k_t$ the week that time point $t$ falls into.
We assumed that the transmissibility of the virus, $\beta_{k_t}$, remains constant between subsequent weeks $[k_t, k_t + 1)$ and employed time-discretization via an Euler-Maruyama approximation of the latent sample path $x$. 
Switching the notation from $t$ to $k_t$, this implies $x_{k_t} \mid x_{k_t - 1}, \sigma \sim \operatorname{N}(x_{k_t - 1}, \sigma^2)$ for the SBM model, which may be viewed as a prior distribution on $x_{k_t}$ and decreases the dimensionality of the stochastic process to $K + 1 < T + 1$. 

Another challenging aspect of parameter estimation is the estimation of the volatility $\sigma$ which is a top-level parameter in the Bayesian hierarchical model; due to the multiple levels of hierarchy in the evidence synthesis model (Figure \ref{fig:functional_relationships}) and the lack of information about it, this difficulty is reflected in the lower effective sample sizes \citep{geyer, stan2024} in all of our analyses. 
Subsequently, this increases the difficulty to break down the transmission rate in \eqref{eq:transmission_matrix_process} into its biological and social components.
Finally, for this multivariate ODE time series model, the solution to \eqref{eq:ODE_system} is approximated numerically via the Trapezoidal rule.
Higher-order numerical ODE solvers can be considered, at the cost of increased computational effort.

The Bayesian hierarchical model is implemented via a dynamic Hamiltonian Monte Carlo algorithm \citep{betancourt} which utilizes gradient information to efficiently explore high-dimensional parameter spaces or complex posterior geometries and to obtain a sample from the posterior distribution of the model parameters given the observed data. 
For a detailed mathematical description of the algorithm, the reader is referred to \cite{chatzilena2}.
The size of the parameter space increases with longer time horizons due to the involvement of the diffusion process in \eqref{eq:Euler_BM} at weekly level, making such a gradient-based MCMC method particularly suitable for the implementation of the Bayesian hierarchical model proposed in the work; simultaneously, this poses a challenge for ODE models, for which computing the gradient is computationally significantly more demanding and subtle than the plain likelihood evaluation \citep{vehtari}. 
The increasing computational cost as a consequence of increasing volumes of data for ODE models is a point raised by \cite{birrell} and \cite{ghosh} and highlights the need for further developments in the area. 
The CPU time required for the MCMC run of each of the SBM and MBM transmission models using Greek and Austrian data is less than 18 hours and is presented in the Supplementary Material, Table S4.

Sampling efficiency was assessed using the effective sample size \citep{vehtari2}, traceplots of model parameters and R-hat. Posterior model assessment was performed by (i) comparing the observed age-stratified mortality counts and the observed mortality counts summed over age groups to simulated samples generated from the posterior predictive distribution of a given transmission model (Supplementary Material, Section S3), (ii) the Deviance information criterion (\citealt{dic}) and cross-validation metrics \citep{psisloo}, see Supplementary Material, Section S3, (iii) assessing the fidelity to the data at a latent level, with visual inspection of the prior and posterior distribution of the contact matrix, (iv) external model validation, comparing the posterior cumulative age-stratified infections in the population to the estimated age-stratified counts of cumulative infections in England, independently estimated via the REACT-2 prevalence survey \citep{ward_react2}.


\section{Data Analysis}\label{section:application}

\subsection{The COVID-19 pandemic data}\label{section:analysis_common_bm}
The proposed methodology is illustrated on freely available data from the COVID-19 pandemic in Greece between September 2020 and March 2021, where $K = 30$ weeks. 
The population was divided in three age groups, $\{0-39, 40-64, 65+\}$. 
During the study period, a national lockdown was implemented in November 2020 until the end of January, 2021. 
Some of the NPIs were relaxed at New Year's Day (Supplementary Material, Table S5). 
In this section we focus on the SBM model which accounts for the inherent variability of contacts in the population as a result of the response of individuals to NPIs which were taken to reduce transmission, expressing the uncertainty in contact structure via prior distributions on the elements of the random contact matrix $C$, which is constrained to not vary in time.

\begin{figure}[ht]
\centering
\includegraphics[scale=0.47]{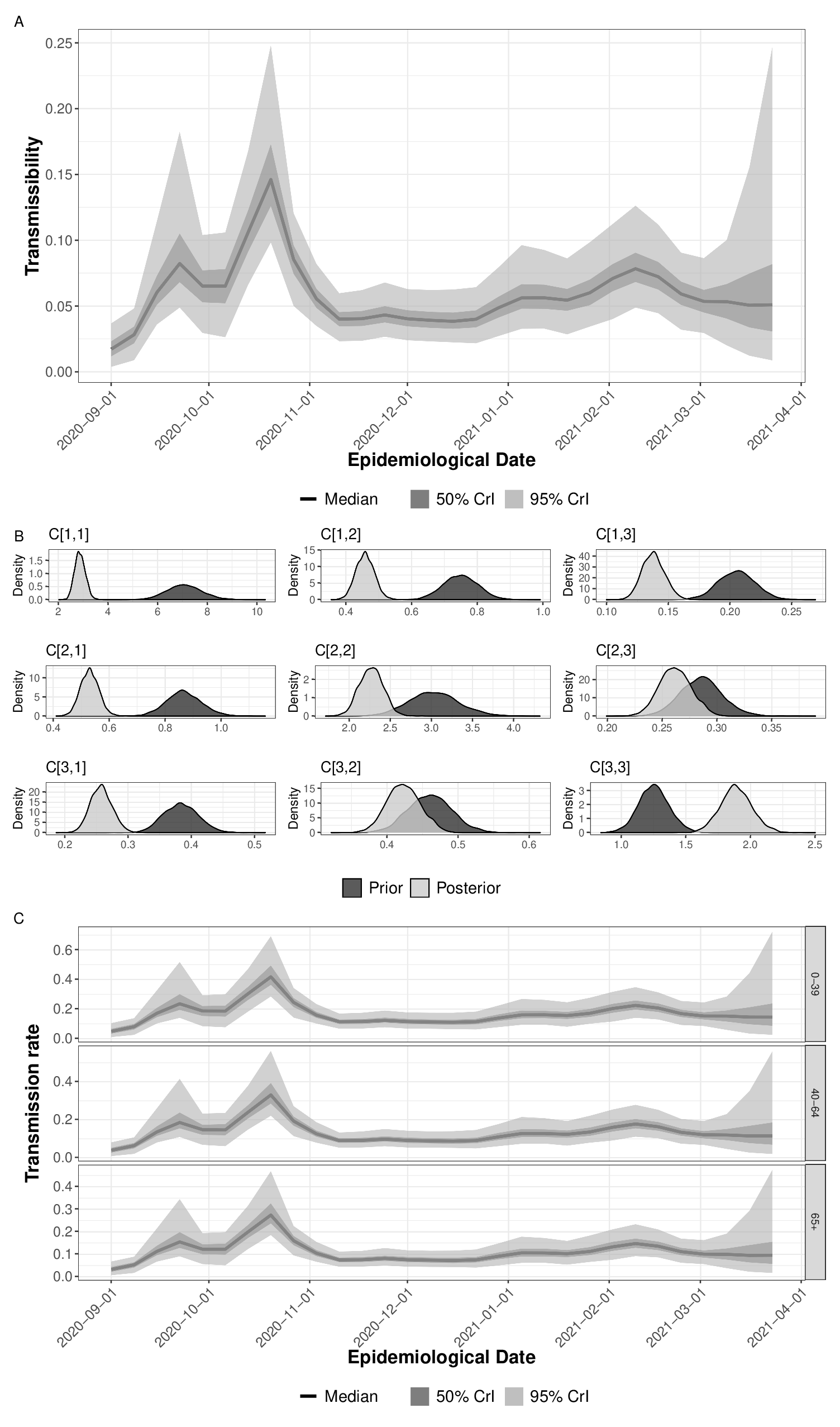}
\caption{Greece - Components of the age-stratified transmission dynamics under the SBM transmission model. Estimated median posterior trajectory (50\% and 95\% credible intervals, CrI) of the transmissibility of SARS-CoV-2 (panel A); 
 prior and posterior distributions of each element of the contact matrix (panel B);
estimated median posterior trajectory (50\% and 95\% credible intervals, CrI) of the age-stratified transmission rate (panel C).
}
\label{fig:common_bm}
\end{figure}

However, the assumption made in \eqref{eq:transm_rate_commom_bm} restricts the age-stratified transmissibility to be expressed by an overall transmissibility for the population, which is dominated by the transmissibility of the age group that drives the transmission of the virus, at the time-points of the observations (Figure \ref{fig:common_bm}, panel A). 
This non-realistic assumption does not properly account for the change in the behaviour of individuals at each age group over time as a result of the implementation of NPIs during the study period.
As previously discussed in Section \ref{section:estimation}, the lack of sufficient information about the volatility $\sigma$ increases the difficulty to break down the transmission rate in \eqref{eq:transm_rate_commom_bm} into its biological and social components, creating an identifiability issue.
Therefore, any differences in the virus transmission procedure across strata are attributed to the random contact matrix elements $C_{\alpha\alpha'}$, justifying the notable differences between the prior and the posterior distribution of the contact matrix (Figure \ref{fig:common_bm}, panel B).
Consequently, the age-stratified transmission rate trajectories (Figure \ref{fig:common_bm}, panel C) only differ in terms of magnitude at the time-points of the observations.

The analysis of Austrian mortality counts revealed the same view regarding the ability of the SBM model to capture the 
age-stratified trends in SARS-CoV-2 transmission (Supplementary Material, Figure S15).
These findings suggest that the SBM model is not flexible enough to accommodate for age-stratified trends in SARS-CoV-2 transmission, motivating the need for a more flexible model which can better estimate the volatility of the diffusion process. For both countries, the observed data are plausible under the posterior predictive distribution of the SBM transmission model. Additionally, no convergence issues were detected (Supplementary Material, Section S2.1).

\subsection{Model expansion}
The assessment of the fidelity of the SBM model to the data at a latent level revealed an issue which may be seen as prior-data conflict at a latent level. 
We resolve this issue by expanding the SBM model to the MBM model in the spirit of \cite{gustafson} and inspecting the effect on the contact survey data. This type of evidence reconciliation is not trivial to assess and it appears that inspecting the prior to posterior updates for the random but time-independent contact matrix yields a reasonable approach to surmount the conflict.

The mortality counts that are available from the national surveillance systems consistently indicate an increasing number of COVID-19 attributable deaths with age; for the case of Greece the respective time series for the younger age group, i.e. $\{0-39\}$, is sparsely-informed, while the elder age groups $\{40-64, 65+\}$ are well-informed; a similar observation can be made for the Austrian dataset (Supplementary Material, Figures S11-S14). 
In the context of diffusion-driven transmission models it would be more sensible to group younger ages together so as to avoid sparsely-informed time series which are more difficult to fit and instead allow for more groupings in elder ages, i.e. by 5- or 10-year intervals, depending on the information that becomes available from the national surveillance systems. 
In such a case, where the number of age groups $A$ is small, the mortality data do not provide much information on the top-level volatilities $\sigma_{\alpha}, \alpha \in \{1,\ldots,A\}$ of the MBM Bayesian hierarchical model, reflecting the difficulty in the estimation of the volatility $\sigma_x$ involved in the SBM model. 
Similarly to the SBM model, posterior inference of the volatilities $\sigma_{\alpha}, \alpha \in \{1,\ldots,A\}$ has resulted in lower effective sample sizes compared to the rest of the MBM model parameters. 

\begin{figure}[ht]
\centering
\includegraphics[scale=0.46]{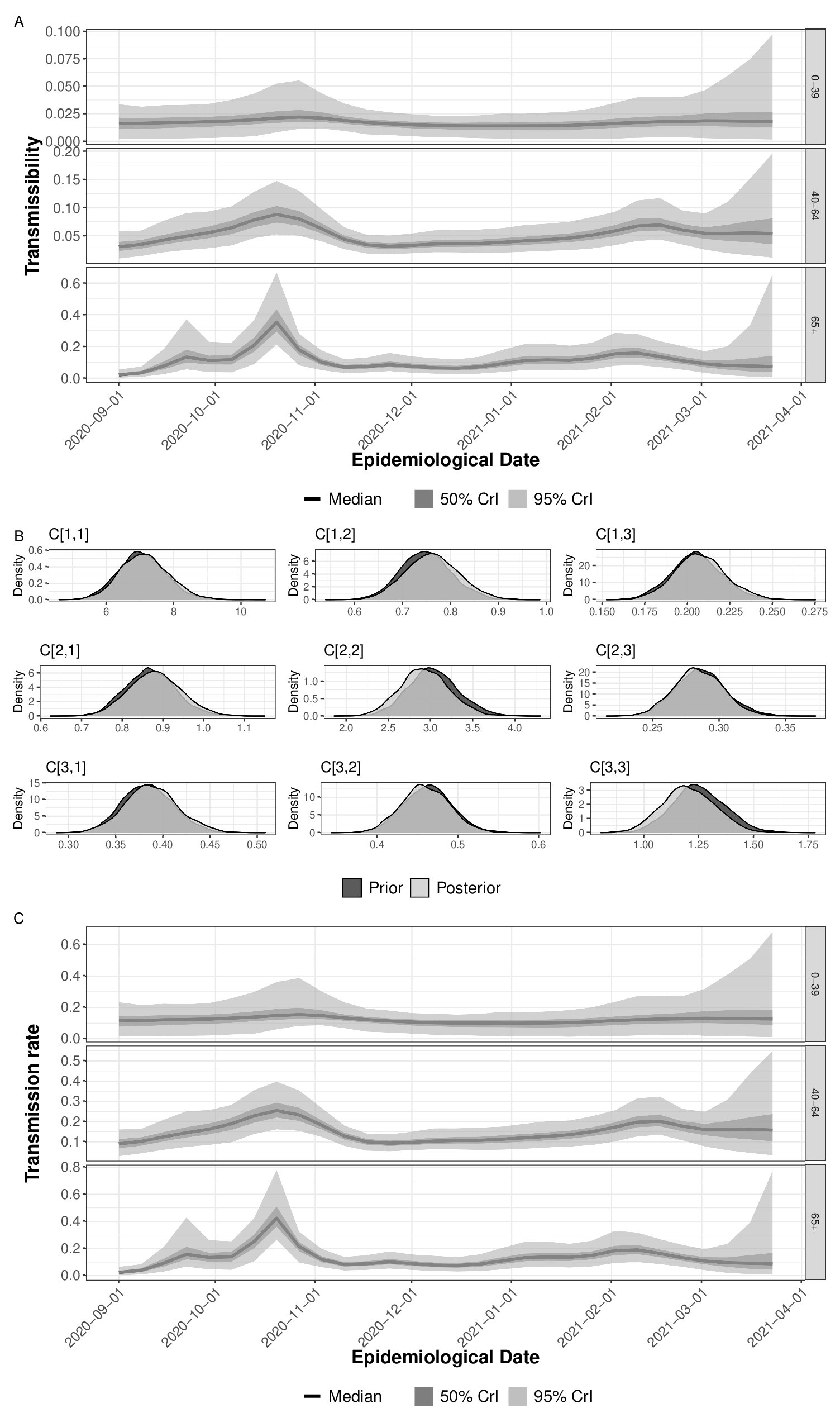}
\caption{Greece - Components of the age-stratified transmission dynamics under the MBM transmission model. Estimated median posterior trajectory (50\% and 95\% credible intervals, CrI) of the transmissibility of SARS-CoV-2 (panel A); 
prior and posterior distributions of each element of the contact matrix (panel B);
estimated median posterior trajectory (50\% and 95\% credible intervals, CrI) of the age-stratified transmission rate (panel C). 
}
\label{fig:indep_bm}
\end{figure}

The assumption of independent diffusions for the biological components of the transmission rate in \eqref{eq:transm_rate_indep_bm} allows for better reconstructing of the age-stratified drivers of transmission (Figure \ref{fig:indep_bm}, panels A and C).
The order of magnitude of the transmission rates changes between the SBM and MBM models because of the change in the order of magnitude in the posterior number of contacts (Figures \ref{fig:common_bm} and \ref{fig:indep_bm}, panel B).
Overall, the expanded transmission model offers better interpretability and more flexibility than the SBM model in inferring dynamic transmission rates and enables the reconstruction of the age-stratified drivers of transmission. 
We discuss a quantitative comparison of the two models using information criteria in Supplementary Material, Section S3. The execution of the MCMC sampling routine for England took 21 hours under the computational setting described in Supplementary Material, Section S2.1. No convergence issues were detected and it appears that the observed data can be considered as a plausible realisation of the posterior density of the MBM transmission model.

The MBM model is validated in the period 2020-03-02 to 2020-09-27 ($K = 30$ weeks) using the estimated age-stratified numbers of cumulative infections in England from the random contact prevalence survey \citep{ward_react2}. 
In the analysis the population was divided in three age groups, $\{0-39, 40-59, 60+\}$. 

\begin{figure}[p]
\centering
\includegraphics[scale=0.158]{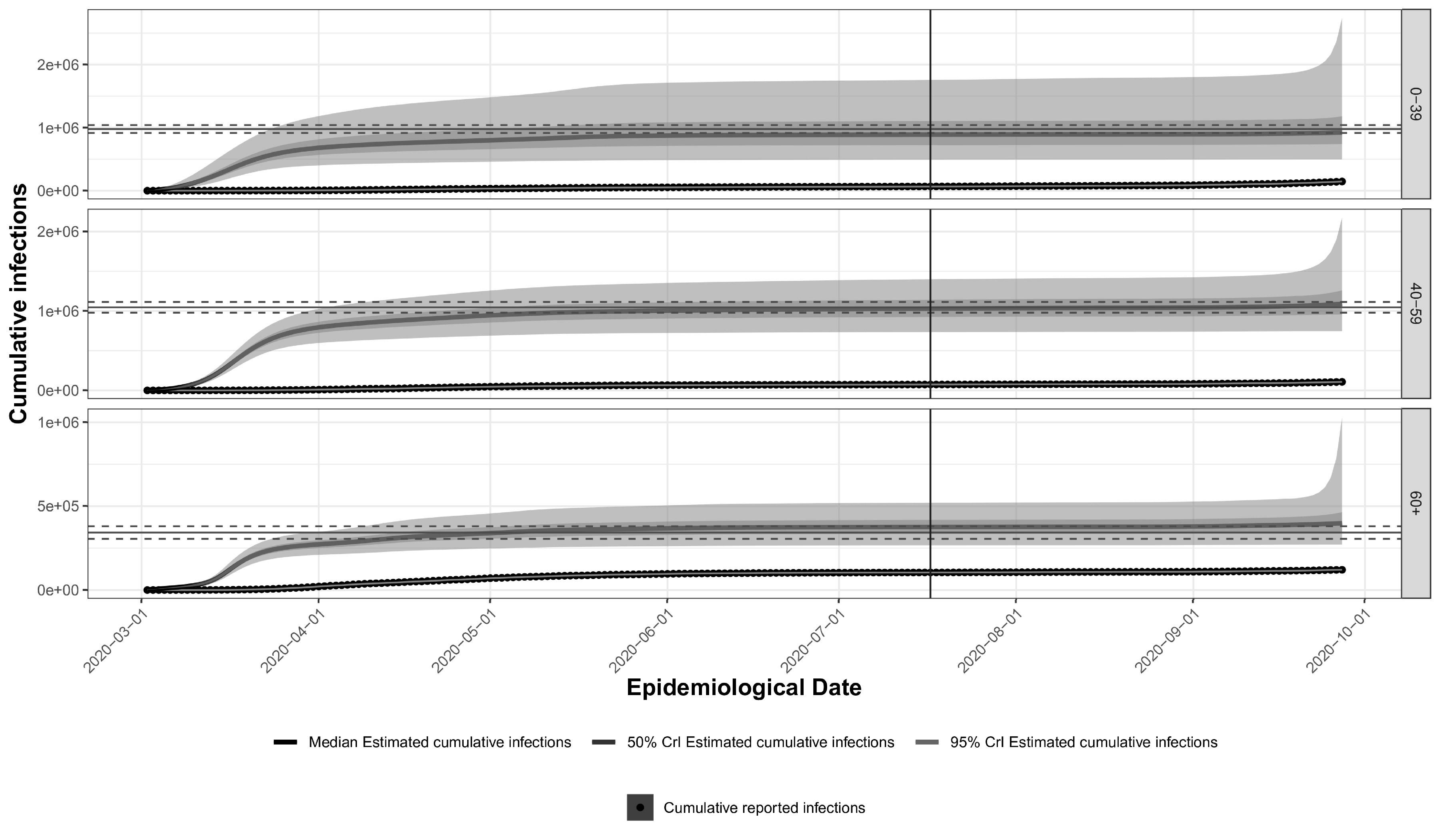}
\caption{External validation - England. Posterior median age-stratified cumulative infections (50\% and 95\% credible intervals, CrI) under the MBM model. The age-adjusted estimated counts and the respective 95\% confidence intervals of cumulative infections based on REACT-2 \citep{ward_react2} are represented by horizontal lines. The vertical line corresponds to middle of July 2020.}
\label{fig:model_validation}
\end{figure}

The estimated counts of cumulative infections from REACT-2 were adjusted by the age distribution of the population  based on the three age groups $\{0-39, 40-59, 60+\}$. 
Figure \ref{fig:model_validation} shows the agreement of the model estimates with the estimates from REACT-2 and demonstrates the ability of the MBM model to capture the level of under-ascertainment of infections over time and by age group.

\subsection{Effectiveness of interventions}
We assess the interventions in transmission (Supplementary Material, Table S5) for Greece based on the findings of the MBM transmission model. The findings for Austria are discussed in the Supplementary Material.

The trajectory of the posterior transmission rate for individuals aged $60+$ (Figure \ref{fig:indep_bm}, panel C) resembles the respective trajectory shown in Figure \ref{fig:common_bm} Panel C, thus stressing the role of this group to the evolution of the epidemic. 
The transmission rate of the virus reached its highest levels in the third week of October 2020; the largest magnitude is demonstrated for the age group $\{60+\}$ (Figure \ref{fig:indep_bm}, panel C), providing evidence to support public health authorities in taking appropriate measures to decrease transmission within and between-groups.
The consecutive decline was steeper for individuals aged $40+$ in the following two weeks. Interventions to limit all indoor/outdoor mass/public gatherings until the end of January 2021 were particularly effective in individuals aged $40+$, keeping virus transmission to low levels, at about 10\%.
The interventions implemented at the end of January 2021 to limit outdoor mass/public gatherings of over 100 participants were successful in controlling the gradual increase in transmission in individuals aged $40+$. Partial relaxation of the measure regarding the closure of day cares, primary and secondary schools in early February 2021 for a month appears to have contributed in controlling transmission.

\section{Discussion}\label{section:discussion}
In this paper we have presented an epidemic model driven by the distinct features of COVID-19 data. We proposed a flexible Bayesian evidence synthesis framework that enables a data-driven approach to inferring the mechanism governing COVID-19 transmission, based on diffusion processes that are a-priori independent. The modularity of the Bayesian approach allows for assessing fidelity to the data at a latent level and resolving the corresponding prior-data conflict via expanding the model to incorporate distinct diffusions for each age group.
\begin{sloppypar}
There are several advantages associated with our approach. Our model is primarily driven by the reported daily COVID-19 attributable mortality counts; a key strength of our Bayesian evidence synthesis framework is that multiple data sources which are publicly available across countries are integrated to provide a robust overall picture of the epidemic nationwide: daily COVID-19 attributable mortality counts, contact surveys and the age distribution of the population. 
This increases the applicability of the model to multiple countries, since these data sources are made available by the national surveillance systems and we do not rely on bespoke surveys. 
We develop data-driven estimation of the transmissibility of the virus via mutually independent diffusion processes thus removing the need for additional information on the timing of specific interventions and hypotheses about their impact on transmission. 
Instead, more general assumptions need to be made, such as on the smoothness of the latent sample path of the age-stratified diffusion that we wish to infer. The Brownian motion is only one of the possible models of our transmission framework since in principle other diffusions can be considered such as the Ornstein-Uhlenbeck process \citep{ornstein}. 
The transmission framework can also be extended to continuous time, thus targeting a hypo-elliptic diffusion, by potentially lowering the time discretisation step and employing a suitable numerical scheme, e.g. \cite{pokern} or \cite{ditlevsen}.
\end{sloppypar}

Our approach allows policymakers to assess the effect of NPIs on each age group.
The model simultaneously estimates multiple metrics of outbreak progression: by modeling mortality counts instead of inferring only from confirmed infections, the Bayesian hierarchical model is more robust to changes in COVID-19 testing policies and the ability of national healthcare systems to detect COVID-19 infections, i.e. through contact tracing, and provides valuable information about the age-stratified transmission rates, latent number of infections, reporting ratio and the effective reproduction number (Supplementary Material, Sections S4-S7), where uncertainty is accounted for naturally via MCMC.

Our modeling approach is subject to limitations. 
Age-stratified counts of vaccinations are not accounted for in the latent transmission epidemic model in the form of covariates, but this does not pose an important limitation for the considered time period, with low counts of administered vaccinations until the end of March 2021. 
In its current form the proposed hierarchical model is better suited to the pre-vaccination era.
In addition, the model is limited by variations in the reporting procedure of deaths and mortality definitions across time and countries (Supplementary Material, Section S1); the dates of daily reported deaths may deviate from the actual dates of deaths. The use of diffusion processes can be viewed as an attempt to model the noise within the observational model and mitigate such issues.
Similar to the Bayesian models of \cite{monod} and \cite{wistuba}, we make parametric assumptions regarding the distribution of the time between infection and reported death in \eqref{eq:itd} (i.e. it is assumed to be the same across age groups and constant during the study period). 
While factors like hospital capacity and vaccine efficacy are expected to affect the infection-to-death distribution, such parametric assumptions might not be generally transferable to countries with different reporting characteristics and healthcare systems. Additionally, reporting delays in the registration of deaths, coupled with the assumption that the time from infection to death is, on average, three weeks, results in the age-stratified death counts being uninformative about patterns of incidence of death over the previous two weeks so that estimates of new infections appear as a lagged indicator \citep{pellis}.

Some further limitations are commonplace in the literature. We model a closed environment with no  infections imported from outside of the population. 
Additionally, our model does not explicitly account for hospital-acquired infections which may have contributed to overall transmission and to persistence of infection in periods of high transmission. 
Age-specific counts of hospitalizations and intensive care unit cases (where publicly available) could help relax the parametric assumptions regarding the infection-to-death distribution and assist in the estimation of the course of the COVID-19 pandemic.
The age-stratified IFR is a crucial component of the observation model as it enables estimation of the expected number of new infections from the observed mortality counts.
An assumption that has been made is that the age-stratified IFRs remain constant during the study period. 
However, these are expected to vary in time and across populations and locations due to a number of factors, such as SARS-CoV-2 variants, vaccine efficacy, the age distribution of the population, the age distribution of infection and the age distribution of mortality which change over time, underlying health conditions, access to medication, the overall burden to the healthcare system and other factors \citep{brazeau}. 
Detailed such data could be incorporated leading to a dynamic age-stratified IFR. 
While the proposed model does not account for the daily laboratory-confirmed COVID-19 infection counts, in principle such an inclusion can be allowed conditionally on the understanding and the appropriate modeling of the biases affecting the data.

In future work we shall focus on two directions to expand our model. The first will aim to incorporate dynamic age-stratified IFR estimates and integrate further healthcare surveillance data like age-stratified counts of vaccinations, hospitalizations and intensive care unit cases where these data are available.
The second extension relates to multi-task Gaussian Processes which have been recently implemented for Bayesian heterogeneously mixing infectious disease models \citep{xu, kypraios}, and it could amend our model via exchangeable diffusion processes, reflecting a-priori beliefs that there is a shared structure between the dynamic transmission rates for individuals of different age groups.

\section*{Acknowledgments}
The authors are grateful to the anonymous referees for their constructive comments that improved the paper. The authors wish to thank Anastasia Chatzilena, Theodore Kypraios and David Rossell for their helpful and constructive comments.

\section*{Funding}
Lampros Bouranis was supported by the European Union's Horizon 2020 research and innovation programme under the Marie Sklodowska-Curie grant agreement No 101027218.

 \noindent{\it Conflict of Interest}: None declared.

\section*{Supplementary Material}

The reader is referred to the online Supplementary Material at\textit{ Journal of the Royal Statistical Society: Series A} for additional information regarding this work.
The \pkg{Bernadette} package \citep{bernadette2, bernadette} for \proglang{R} \citep{rsoft} implements the methodology in this paper. 
The \proglang{R} code and documentation to reproduce the analysis are available at \url{https://github.com/bernadette-eu/indepgbm}.

\bibliographystyle{apalike}
\bibliography{indepbm}


\pagebreak

\pagebreak

\pagebreak

\pagebreak

\pagebreak

\end{document}